\documentclass[twocolumn]{aastex631}

\usepackage{amsmath}
\usepackage{amsfonts}
\usepackage{graphicx}
\usepackage{hyperref}
\usepackage{float}
\usepackage{multirow}
\usepackage[mathlines]{lineno}
\usepackage{subfigure}
\begin{document}

\title{Introducing RAFIKI: Refining AGN Feedback in Kinetic Implementations}

\author{Skylar Grayson}
\affiliation{School of Earth and Space Exploration, Arizona State University, P.O. Box 876004, Tempe, AZ 85287, USA}

\author{Evan Scannapieco}
\affiliation{School of Earth and Space Exploration, Arizona State University, P.O. Box 876004, Tempe, AZ 85287, USA}

\author{Romeel Dav\' e}
\affiliation{Institute for Astronomy, Royal Observatory, University of Edinburgh, Edinburgh EH9 3HJ, UK}
\affiliation{Department of Physics and Astronomy, University of the Western Cape, Bellville, Cape Town 7535, South Africa}
\affiliation{Stellenbosch Institute for Advanced Studies, Mosterdrift, Stellenbosch 7600, South Africa}

\author{Arif Babul}
\affiliation{Department of Physics and Astronomy, University of Victoria, Victoria BC V8P 5C2, Canada}
\affiliation{Institute for Astronomy, Royal Observatory, University of Edinburgh, Edinburgh EH9 3HJ, UK}
\altaffiliation{Leverhulme Visiting Professor}

\author{Renier T. Hough}
\affiliation{Centre for Space Research, North-West University, Potchefstroom 2520, South Africa}
\affiliation{National Institute for Theoretical and Computational Sciences (NITheCS), Potchefstroom 2520, South Africa}

\newcommand{\ES}[1]{\textcolor{blue}{#1}}

\submitjournal{The Astrophysical Journal}

\begin{abstract}

Modern cosmological simulations have matured to the point of reproducing realistic galaxy populations across cosmic time. Yet these simulations rely on feedback from active galactic nuclei (AGN) to quench massive galaxies, and the details of this process remain poorly understood. To address this issue, we introduce RAFIKI (Refining AGN Feedback In Kinetic Implementations), a novel suite of simulations built upon SIMBA-C that vary the energetic efficiencies of jets and winds driven by AGN. Unlike the fiducial SIMBA-C simulation, RAFIKI separates the efficiencies of these two feedback modes, enabling a detailed study of their impact on galaxies, black holes, and the circumgalactic medium. We find that even with enhanced mass loading, the lower-velocity quasar-type wind mode cannot quench massive galaxies. However, it plays a significant role in regulating black hole growth and star formation in intermediate-mass galaxies. We also uncover degeneracies in the parameter space that highlight the lack of current constraints on AGN feedback. RAFIKI provides a controlled framework to disentangle these degeneracies using current and upcoming observations. 

\end{abstract}


\section{Introduction}\label{sec:intro}
Cosmological hydrodynamic simulations are an essential component of modern astrophysics, providing a tool for interpreting a  wide range of multiwavelength observations. These models strive to realistically emulate the processes governing galaxy evolution, including gravity and structure formation, gas heating and cooling, chemical evolution, star formation, and feedback from stellar winds, supernovae, and active galactic nuclei (AGN) \citep[see][for a review]{crain2023}. AGN feedback, in particular,  plays a key role, as it is essential to reproduce both the quenching of massive galaxies and their observed stellar mass distribution \citep[e.g.][]{Scannapieco2004,thacker2006,Sjacki2007,Schaye2015,pillepich2018,dave2019}. 

However, substantial uncertainty remains regarding the physics of accretion and outflows from supermassive black holes. The complexity of magnetohydrodynamic plasma in accretion disks is not fully understood, nor are the mechanisms that power outflows \citep[e.g.][]{yuan2014, pathak2025}. The energetics of these outflows are also often unconstrained, and there is still no consensus on how AGN feedback acts to quench galaxies \citep{fabian2012, delucia2026,king2015}. 

Beyond these uncertainties, the low resolution of cosmological simulations means AGN feedback must be treated as `sub-grid' process, which cannot be modeled from first principles and must be approximated on resolved scales. Numerous free parameters are involved in this approximation, and every simulation adopts different prescriptions, sometimes leading to drastically different implementations. Despite this, most modern cosmological simulations are generally successful in matching the stellar properties of galaxies, because the stellar mass function, the stellar-halo mass relationship, and the stellar-black hole mass relationship are often used to calibrate the feedback parameters \citep{crain2023}. 

However, the differences in feedback models become far more consequential when considering diffuse baryons. Gas inflow and outflow rates vary between simulations, sometimes by a factor of 3 or more \citep{crain2023}, and this can impact the distribution of gas on large scales. Baryon fractions, particularly in lower-mass halos, are not well constrained observationally and differ drastically between simulations \citep{davies2020}, and there is growing evidence that most simulations do not align with observations of gas reservoirs. Studies comparing stacked observations of the hot phase of the circumgalactic medium (CGM) using X-ray emission or the thermal Sunyaev Zel'dovich effect suggest that simulations such as SIMBA and EAGLE are failing to capture all the processes shaping the CGM \citep[e.g.][]{grayson2023, grayson2025}. Thus there is a growing need to better understand and constrain the physical drivers of AGN feedback.

Observationally, AGN have historically been separated into two modes, often labeled `radiative' and `kinetic' \citep[see][for reviews]{fabian2012, heckman2014}. The first is thought to occur at high Eddington ratios, where the energy output is primarily through radiation (i.e. quasars), while the second is dominated by kinetic energy input, with accretion disks fueling jets of material that can form large radio lobes far beyond the disk of the host galaxy. These observational trends motivate how AGN feedback is implemented in simulations to various degrees. 

Some simulations, like EAGLE \citep{Schaye2015} and ROMULUS \citep{tremmel2019}, deposit AGN feedback as thermal energy around the black hole, using a single mode with an energy that scales with the black hole accretion rate. The recent OBSIDIAN simulation implements three modes of feedback, wherein very low accretion rates simultaneously power both a jet and a wind driven by the accretion flow \citep{rennehan2024}. Others, like IllustrisTNG \citep{pillepich2018}, Horizon-AGN \citep{dubois2012}, and SIMBA \citep{dave2019} attempt to capture a separated dual-mode behavior, providing two pathways for AGN feedback to occur depending on the accretion rate of the central supermassive black hole. In the case of IllustrisTNG, at high accretion rates the energy is deposited thermally, but at low accretion rates, outward velocity kicks are given to gas particles around the black hole \citep{nelson2019}. 

SIMBA also follows this dual mode prescription, but it only uses kinetic feedback, with high velocities at low accretion rates (the `jet' mode) and lower velocities at high accretion rates (the `wind' mode). The launched outflows are briefly decoupled from the hydrodynamics to avoid numerical inaccuracies, preventing the feedback energy from being entrained in the ISM near the black hole such that the energy of the jet mode is deposited on scales comparable with observations of radio lobes. SIMBA also stands out in its choice of directionality, as the velocity kicks are not random, as in IllustrisTNG and OBSIDIAN, but rather the gas is ejected in a bipolar fashion orthogonal to the disk. When jets are at full speed in relatively gas-poor galaxies ($M_{\rm gas}/M_*<0.2$), SIMBA also includes an `X-ray heating' mode, which injects a small percentage of the jet energy as thermal energy around the black hole \citep{dave2019}.

A common trend in these and other simulations that use dual mode AGN feedback is the dominance of one mode over the other in shaping galaxy properties. In both SIMBA and Illustris-TNG, the high velocity kinetic mode associated with low accretion rates is most important for quenching star formation and shaping a realistic galaxy population at high mass scales \citep{Scharre2024, tillman2025}. It is also the mode that impacts the large-scale environments, changing the properties of the circumgalactic medium and impacting the dynamics of the intracluster medium. By contrast, the high accretion rate mode plays a less significant role, although it does have an impact at lower mass scales. In SIMBA, this so-called `wind' mode is believed to play a small role in quenching intermediate mass galaxies, with $M_* \approx 10^{9.5}$ to $10^{10} M_\odot$ \citep{Scharre2024}. 

This difference in impact is a natural consequence of the design of SIMBA's feedback model, as the wind mode is much less energetic than the jets. However, it is not clear observationally that this should be the case, as the energetics of AGN are still largely unconstrained \citep{moe2009, fabian2012, chamberlain2015}. It is possible that the quasar-like mode could play a more important role in quenching galaxies and impacting the CGM in ways that can be observable by current and upcoming surveys.   

In this work, we present a novel suite of simulations named RAFIKI (Refining AGN Feedback In Kinetic Implementations) that modifies the AGN feedback treatment in SIMBA-C (a modified version of SIMBA that includes more realistic chemical enrichment modelling \citep{Hough2023}). These runs maintain all other facets of SIMBA-C, but modify the AGN feedback such that the mass loading of the two kinetic modes can be turned up or down independently, allowing for a more complete examination of the impact of lower-velocity winds driven by AGN activity. In this paper, we use RAFIKI to answer the question: can a lower velocity wind model successfully quench massive galaxies? We also explore in detail the impact each feedback mode has on stellar, black hole, and baryon properties. 

The structure of this work is as follows. In Section \ref{sec1} we explain the simulation methodology, including an overview of SIMBA and SIMBA-C (\ref{sec1_1}), a discussion of the modifications we made to generate the RAFIKI suite (\ref{sec1_2}), and an overview of the RAFIKI data products (\ref{sec1_3}). In Section \ref{sec2} we present galaxy properties and scaling relations for the RAFIKI runs, and in Section \ref{sec3} we explore the implications of these results in the context of AGN modeling and the impact of galactic winds on galaxy evolution.

\section{Simulation Methodology}\label{sec1}

\subsection{SIMBA and SIMBA-C}\label{sec1_1}

In this work, we modify the simulation SIMBA-C \citep{Hough2023}, which updates the chemical enrichment models in SIMBA \citep{dave2019}. Here we provide a brief overview of the simulation, before going into detail on the treatment of AGN feedback and how we have modified it.

SIMBA-C is based on the hydrodynamic and gravity solver GIZMO, which uses a meshless-finite mass methodology while preserving the mass of fluid elements, and handles shocks via a Riemann solver with no artificial viscosity \citep{springel2005, hopkins2015}. The GRACKLE-3.1 library is used to model radiative cooling and photoionization heating, including self-consistent self-shielding assuming an attenuation of the ionizing background dependent on the gas density \citep{smith2017,rahmati2013,haardt12}. The code includes feedback from core-collapse supernovae, Type Ia supernovae, and asymptotic giant branch stars, which play a large role in setting properties of lower-mass galaxies \citep{Scannapieco2000,Scannapieco2001,Thacker2002,Springel2003}. These stellar-feedback driven winds are launched  with mass and momentum scalings based on the FIRE simulations \citep{aa2017}. SIMBA-C updates the chemical enrichment model in SIMBA to track 34 elements (H to Ge), and it uses updated metal injection models during stellar feedback events to align with recent constraints on supernova yields \citep{kobayashi2020}.  The SFR in SIMBA-C is dependent on the H$_2$ density following 
\begin{equation}
\text{SFR} = \frac{\epsilon_* \rho_{\rm H_2}}{t_{\text{dyn}}},
\end{equation}
where $t_{\text{dyn}}$ is the dynamical time, and the star formation parameter $\epsilon_*$=0.026.

\subsection{Black Holes and AGN Feedback}\label{sec1_2}

In SIMBA-C, black holes are seeded at a mass of $10^4 M_\odot$ in galaxies with a stellar mass $> 6 \times 10^8 M_\odot$. Black holes grow using a two-mode accretion model, with cold gas following a torque-limited accretion model described in \cite{aa2017}, and hot gas following Bondi-Hoyle-Lyttleton accretion. The former is capped at three times the Eddington accretion rate, while the latter cannot exceed the Eddington rate.

This accretion drives AGN feedback, with SIMBA-C using two main feedback modes referred to below as `winds' and `jets'. It also includes `X-ray feedback', meant to input energy associated with X-rays coming from the accretion disk. This mode is only activated during the jet mode and when $M_{\rm gas}/M_*<0.2$. The X-ray heating has been shown to have a minimal effect on the galaxy stellar mass function (GSMF) and has a very small role in the overall energy budget of the AGN \citep{dave2019}. For this reason, in RAFIKI we remove the X-ray feedback, and focus only on the kinetic implementations. 

The two kinetic modes are motivated by observations of outflows that generally fall into two categories dependent on the Eddington ratio ($f_{\rm Edd} = \dot{M}_{\rm BH}/\dot{M}_{\rm Edd}$, where $\dot{M}_{\rm BH}$ is the accretion rate of the black hole) \citep{heckman2014}. At high $f_{\rm Edd}$ AGN energy is dominated by electromagnetic radiation that can drive outflows at velocities generally around $\approx$ 1000 km s$^{-1}$.  There is evidence of higher-velocity outflows in radiation-dominated AGN \citep[i.e.][]{Zakamska2016,fiore2017, xu2025}, but the SIMBA-C model follows the 500-1000 km s$^{-1}$ velocities found in \cite{perna2017}.  At low Eddington ratios ($f_{\rm Edd} <$ a few percent) we find objects with energetic output dominated by kinetic energy transported by collimated jets with velocities up to the order of 10000 km s$^{-1}$.  SIMBA-C divides feedback into two regimes based on the Eddington ratio. When $f_{\rm Edd} > 0.2$, the wind mode is active, and when $f_{\rm Edd} < 0.2$, the jet mode is active. Beyond the $f_{\rm Edd}$ criterion, the jet mode is enabled only once a black hole exceeds a randomly assigned threshold mass between $7 \times 10^7\,M_\odot$ and $1 \times 10^8\,M_\odot$. This prevents low-mass black holes with temporarily low accretion rates from powering jets. In RAFIKI, we simplify this approach, setting a fixed mass threshold of $4 \times 10^7 M_\odot$ for jets to turn on, which was the minimum mass set in the fiducial SIMBA \citep{dave2019}.  

The opening angle for both modes is set at zero in SIMBA-C, such that particles are ejected in a purely bipolar direction perpendicular to the angular momentum vector of the inner disk (generally defined as the 256 nearest gas particles to the black hole, see \cite{aa2017} for more details). In RAFIKI, we maintain this treatment for the jets, but modify the radiative wind mode such that particles are ejected randomly in an isotropic manner. We find this to be more realistic, as the motivation of bipolar ejection is based on observations of radio galaxies with strong jets, and not the quasar-type feedback meant to be emulated by the wind mode of this model \citep[e.g.][]{elvis2000}. This change will also allow for broader extensions of the results of this study to the myriad of other models that implement isotropic quasar-like feedback \citep[e.g][]{thacker2006,Schaye2015,weinberger2017,pillepich2018}. 

For the wind mode, the velocities of the ejected particles are motivated by SDSS observations of X-ray selected AGN in \cite{perna2017}, set by 
\begin{equation}
v_w = 500+ 500\text{log}_{10} \left(\frac{M_{\rm BH}}{10^6 M_\odot}\right)^{1/3} \text{km s}^{-1},
\end{equation}
where $M_{\rm BH}$ is the mass of the black hole. When the jet mode is activated, the velocities are set by 
\begin{equation}
v_j = v_w + 7000 \text{log}_{10} \left( \frac{0.2}{{\rm Max}(f_{\rm Edd}, 0.2)} \right) \text{km s}^{-1}. 
\end{equation}

The mass loading factor for the ejected particles for both modes is motivated by observations of molecular outflows \citep{king2015} and is parameterized by the momentum flux
\begin{equation}\label{og_p}
\dot{P} = 20L/c,
\end{equation}
where $c$ is the speed of light and $L$ is the bolometric luminosity of the AGN, and is defined as 

\begin{equation}
L = \eta \dot{M}_{\rm BH}c^2.
\end{equation}
Here, $\dot{M}_{\rm BH}$ is the accretion rate of the black hole, and $\eta$ is the radiative efficiency, set at 0.1 in SIMBA-C. From these equations, we find that the mass loading of the wind is dependent on the black hole growth rate and the velocity of the particle, $v$, following:
\begin{equation}
\dot{M}_{\rm out} = \frac{20 \eta c \dot{M}_{\rm BH}}{v}.
\end{equation}

Here we can see how the structure of SIMBA's feedback model leads to the wind mode being unable to quench massive galaxies. The feedback is designed such that the momentum input is constant and the energy scales with velocity. As the mass of halos increases, the velocity in the wind mode does not increase at a fast enough rate to keep heating the halo gas to the virial temperature, and the mass loading is too low to counteract this effect. Therefore the higher velocities present in the jets are needed to heat gas sufficiently to quench galaxies in large halos. It is also worth noting that outflow momentum fluxes are highly uncertain and vary significantly between observations of molecular outflows \citep[e.g.][]{Faucher2012, king2015, Fluetsch2019, Spilker2025}, suggesting that other methods of parameterizing feedback models might better capture the range of outflow properties seen observationally.

This motivates the largest difference between SIMBA-C and RAFIKI. In RAFIKI, we parameterize the feedback based on an energetic efficiency, such that the kinetic power of the AGN outflow is given by

\begin{equation}
\dot{E}_{\rm kin} = \epsilon \eta \dot{M}_{\rm BH}c^2,
\end{equation}

where the efficiency $\epsilon$ corresponds to how much of the bolometric luminosity goes into driving the winds. We keep the same velocity prescription as SIMBA-C, and the momentum flux is dependent on the velocity following
\begin{equation}\label{new_p}
\dot{P} = 2\epsilon L/v. 
\end{equation}
Crucially, we set $\epsilon$ separately for the wind and jet modes, hereafter $\epsilon_w$ and $\epsilon_j$. As the jet mode has been known to be the dominating factor in SIMBA and SIMBA-C, by adding separate processes for the mass loading we can better explore the impacts of the low velocity winds implemented in SIMBA. This distinction is also physically motivated: while the winds may be radiatively powered, the mechanisms that launch jets are fundamentally different, so separating the kinetic power loading of the modes allows a more realistic treatment of the underlying physics. By comparing Equation (\ref{og_p}) and (\ref{new_p}) we can estimate the values of $\epsilon_w$ and $\epsilon_j$ corresponding to the fiducial SIMBA-C treatment. While these values vary based on accretion rates and black hole mass, $\epsilon_w$ could reach a maximum value of $\approx 0.013$, with $\epsilon_j$ nearing 0.2.

\begin{figure*}[htbp]
    \centering
	
	\includegraphics[width = \linewidth]{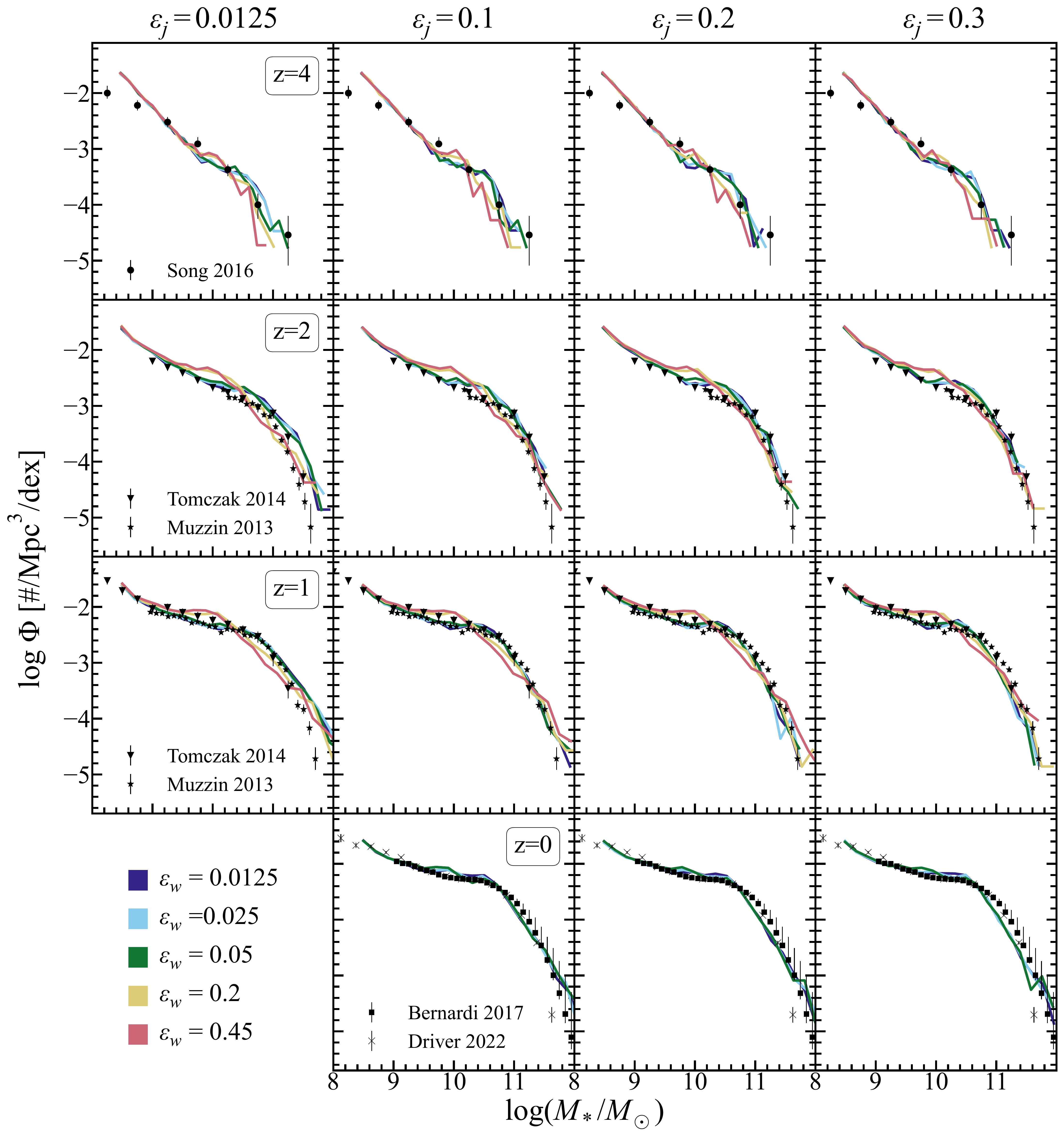}
\caption{ Galaxy stellar mass functions at redshifts 4, 2, 1, and 0 (top to bottom). Each column represents a different value for $\epsilon_j$, with jet mass loadings at 1.125, 10, 20, and 30 $\%$ from left to right. In each panel, RAFIKI runs are compared against observational data in black. At $z$=4, observational points are from \cite{song2016}, at $z$=2 and $z$=1 observational points are from \cite{tomczak14} and \cite{muzzin13} and at $z=0$ observations are from \cite{bernardi2017} and \cite{driver2022}. Runs with $\epsilon_j$=0.0125 were unable to match the high end of the galaxy stellar mass function and were thus not continued past z=1. High  $\epsilon_w$ runs overpredict galaxies at log($M_*/M_\odot$) $\approx$ 10 and underpredict log($M_*/M_\odot$) $\approx$ 11, so these runs were also not continued past z=1. Other parameter choices were all able to successfully reproduce the observed galaxy stellar mass function across cosmic time. }
    \label{fig:gsmf}
\end{figure*}

The RAFIKI suite is not the only recent work that has attempted to understand the impacts of AGN by varying feedback strengths in SIMBA. The Cosmology and Astrophysics with MachinE Learning Simulations (CAMELS) consists of over 4000 runs varying the parameters of the IllustrisTNG and SIMBA subgrid physics \citep{camels2021}. In the SIMBA runs, AGN feedback is parameterized by the speed of the jet mode and the momentum flux of both the jet and wind modes. This has allowed for extensive work detailing how these feedback parameters impact galaxy, gas, and cosmological properties \cite[e.g.][]{moser2022, delgado2023, lau2025}. However, because there is only one parameter setting the momentum flux for both the jet and wind mode, the CAMELS simulation suite is constrained by the same mass-loading formulation as SIMBA.

\subsection{RAFIKI Runs and Analysis}\label{sec1_3}

The RAFIKI suite consists of twenty runs that vary $\epsilon_w$ and $\epsilon_j$. The choices for these values are not all realistic, but rather attempt to probe the effects of adjusting the kinetic power loading in kinetic AGN feedback models while maintaining the energy budget. All runs were carried out in a 50$h^{-1}$ Mpc volume with $512^3$ gas and $512^3$ dark matter particles.  We began the runs at $z$=249 and used a $\Lambda$CDM cosmology following Planck Collaboration VI, with $\Omega_m =0.3$, $\Omega_\Lambda=0.7$, $\Omega_b=0.048$, and $H_0$ = 68 km s$^{-1}$ Mpc$^{-1}$ \citep{planck}. Every model was run down to at least $z=1$. A subset of those that successfully matched the $z=1$ galaxy stellar mass function were ran all the way to $z=0$ (6/20 runs).  

For each run, we generated a galaxy and halo catalog using the yt-based package CAESAR \footnote{Information about Caesar can be found at \url{https://caesar.readthedocs.io/en/latest/}}. The code identifies galaxies using a 6-D friends-of-friends algorithm with a linking length of 0.0056 times the mean interparticle spacing. For each of our runs, CAESAR calculated physical properties of galaxies, generated particle lists for each galaxy and halo, and linked galaxies to their halos while identifying centrals and satellites.

\section{Results}\label{sec2}

\subsection{Stellar Properties}
We begin by exploring the impact of varying feedback models on stellar properties of the galaxies, including the galaxy stellar mass function, the global star formation rate history and the stellar mass-star formation rate relation.

\begin{figure}
\includegraphics[width=\linewidth]{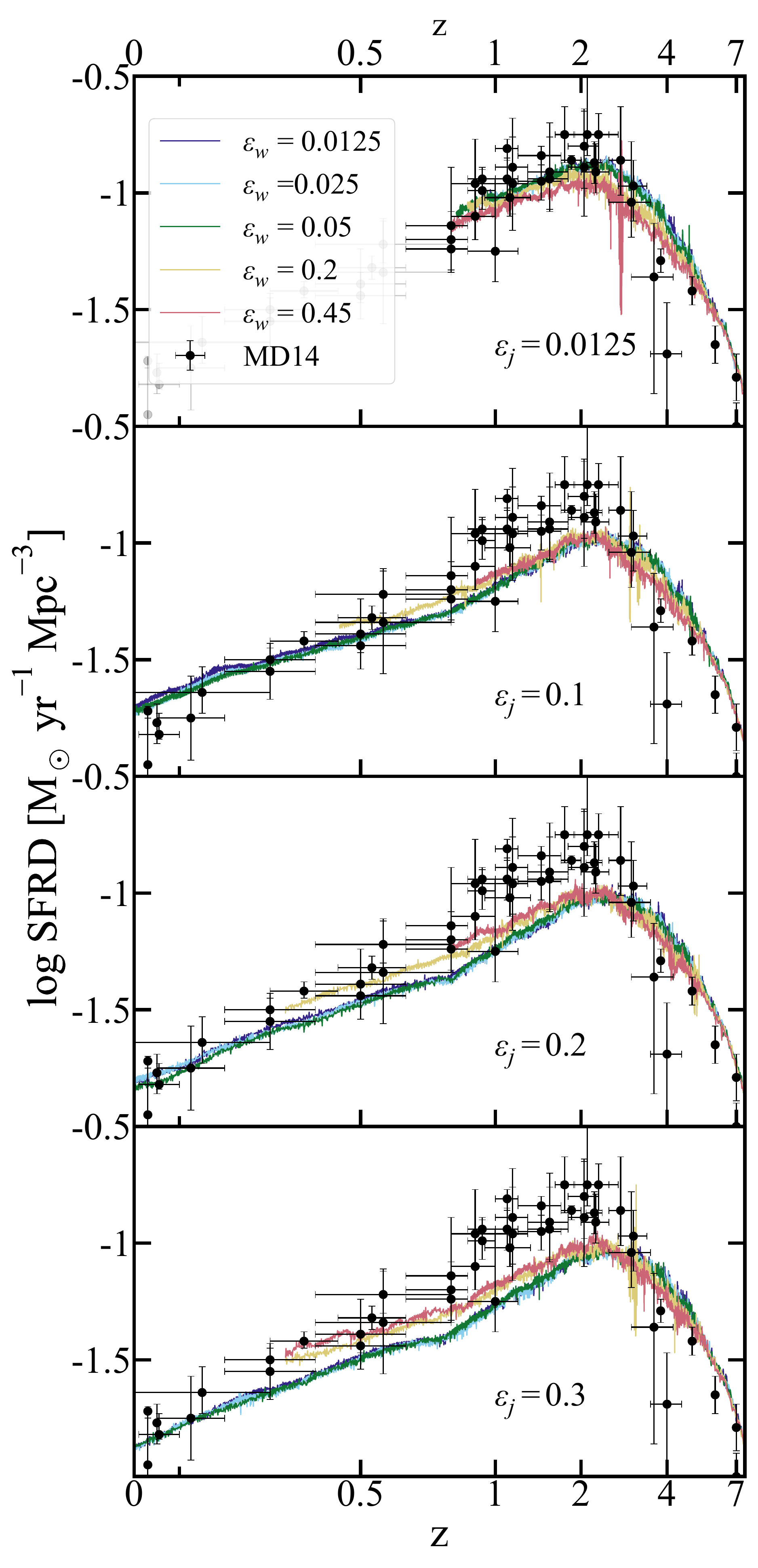}
\caption{ Star formation rate density for all RAFIKI runs. From top to bottom, panels show runs with $\epsilon_j =$ 0.0125, 0.1, 0.2, and 0.3. Black points show observational data from \cite{madau_dickinson}. Higher $\epsilon_j$ values result in a peak in SFRD at earlier times and a slight underprediction of SFRD at low redshift.}
\label{fig:sfrd}
\end{figure}

\begin{figure}
\includegraphics[width=\linewidth]{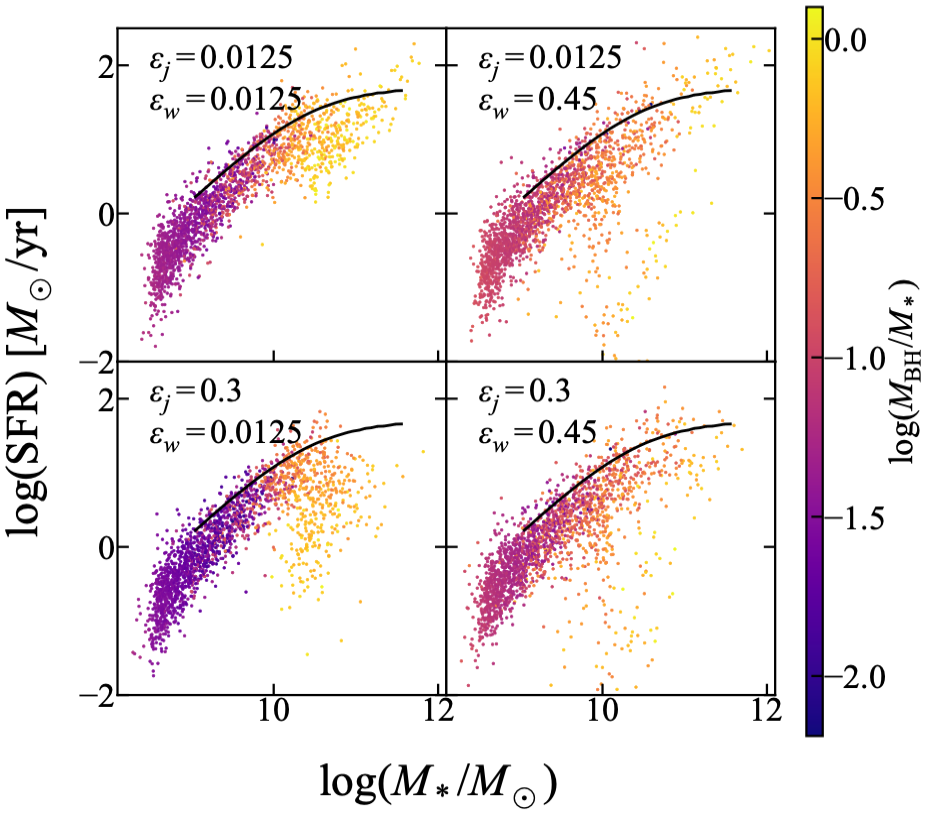}
\caption{Stellar mass versus star formation rate dfor four selected RAFIKI runs at $z=1$ spanning the parameter space, colored by the black hole-stellar mass ratio. Top row shows runs with $\epsilon_j = 0.0125$, bottom row has $\epsilon_j = 0.3$, left column has $\epsilon_w = 0.0125$, and right column has $\epsilon_w = 0.45$. The black lines show the best fit to observational data from \cite{popesso_2023}. The RAFIKI runs consistently underpredict the SFR, which is a trend seen across cosmological simulations.}
\label{fig:main_sequence}
\end{figure}

\subsubsection{Galaxy Stellar Mass Function}

The galaxy stellar mass function (GSMF) tracks the efficiency of halos in converting their baryons to stars. The shape of the GSMF is well measured across cosmic time, and thus provides a strong test of cosmological simulations. The fiducial SIMBA was primarily calibrated using the $z=0$ GSMF, and we have used the $z=1$ GSMF to determine which RAFIKI runs adequately quench massive galaxies and thus should be run to lower z \citep{dave2019}. Figure \ref{fig:gsmf} shows the GSMF for all RAFIKI runs at $z$ = 4, 2, and 1, and 0. The columns represent different $\epsilon_j$ values, with each color representing  $\epsilon_w.$ 

\begin{figure*}
\includegraphics[width=\linewidth]{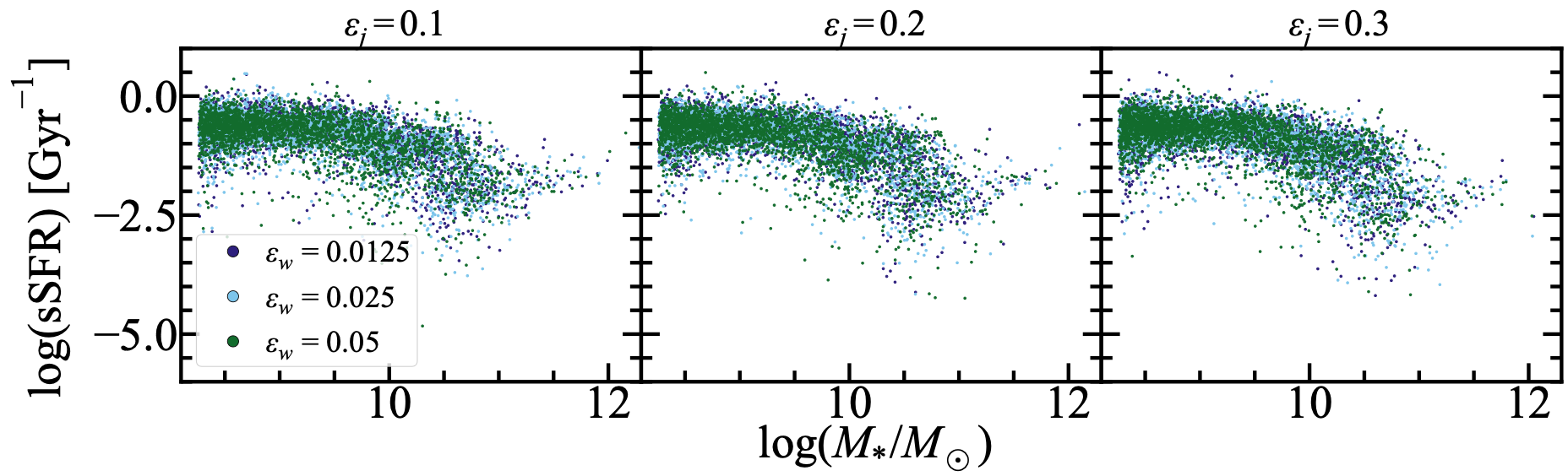}
\caption{Stellar mass against sSFR for RAFIKI runs at $z=0.$ The RAFIKI runs do not include the X-ray heating mode incorporated in SIMBA, which was shown to be important for quenching the most massive galaxies. Here we see that some runs might struggle to quench systems with log($M_*/M_\odot) \approx 12$, as there are some high-mass systems with high sSFR. However, higher $\epsilon_j$ values are more effective, which sSFRs 0.5 dex lower for $\epsilon_j$=0.3 compared to $\epsilon_j$=0.1. }
\label{fig:z0_stellar_sfr}
\end{figure*}

In looking at the leftmost column, containing runs with $\epsilon_j = 0.0125$, we see that with very low efficiencies in the jets we cannot match the high-mass end of the GSMF. Even with the winds experiencing signficantly higher mass loading, there is almost no change in the highest mass galaxy population compared to very low wind efficiencies. It is for this reason that none of these runs were continued past $z=1$. However, values of $\epsilon_j$ from 0.1-0.3 successfully match the high mass function, again with very little change between $\epsilon_w$ values. 

Significant increases to $\epsilon_w$ do have a large effect at lower masses, particularly around log($M_*/M_\odot) \approx 10$. Here, when wind efficiencies are at 20 and 45$\%$ we see a surplus of galaxies, independent of $\epsilon_j$. This then leads to fewer galaxies around  log($M_*/M_\odot) \approx 11$, before a slight surplus past log($M_*/M_\odot) > 11.5$. The effects of the wind mode on intermediate mass galaxies will be explored in more detail in later sections.

\subsubsection{Star Formation Rate Density}

Figure \ref{fig:sfrd} shows the cosmic star formation rate density (SFRD) for all RAFIKI runs. This is  a crucial test for galaxy evolution models. Each panel in Figure \ref{fig:sfrd} represents different values for $\epsilon_j$.  We see in the bottom panels, which show runs with higher $\epsilon_j$, that the peak in SFRD happens at slightly earlier epochs and with slightly lower values than the observational data compiled in \cite{madau_dickinson}. This is consistent with the fiducial SIMBA simulation, which lies below the observed star-forming main sequence \citep{dave2019}. In contrast, the top panel, representing $\epsilon_j=0.0125$, shows a noticeably better agreement with the data, with the peak more closely aligned with observations. This suggests that jet feedback is the dominant factor responsible for lowering the main sequence in the simulation.

Figure \ref{fig:sfrd} also reveals a tension between the two metrics, with the GSMF favoring stronger jets, and the SFRD favoring weaker jets. To reduce the computational expense, we chose to only continue runs that matched the GSMF well. Consequently, we lack data on how successful the $\epsilon_j=0.0125$ runs would be at matching the SFRD at late times, as it is possible that there would be more inconsistencies at low-$z.$ These results highlight that simulations can still struggle to replicate all the global galaxy properties, and the choice of calibration parameters can have a substantial impact.

\subsubsection{Star Formation - Stellar Mass Relationship}

Figure \ref{fig:main_sequence} shows a more detailed look at the star-forming main sequence for four RAFIKI runs representing the extremes of the parameter space at $z=1.$ Prior to the significant quenching of massive galaxies at late epochs, the relationship between the star formation rate (SFR) and stellar mass follows a star-forming main sequence. The points in Figure \ref{fig:main_sequence} are color coded to represent the black hole-stellar mass ratio. In general, all the RAFIKI runs shown underpredict the star formation rate (SFR) relative to compiled observations from \cite{popesso_2023}, although there is a slightly better fit at higher stellar mass. This is a common trend in cosmological simulations \citep{Somerville2015,dave2019}, which generally underpredict the SFR around cosmic noon. 

As was seen in SIMBA, the galaxies that are beginning to be quenched at this epoch are ones with a higher $M_{\rm BH}/M_*$ ratio. In general, runs with a high $\epsilon_w$ have lower $M_{\rm BH}/M_*$ ratios at high stellar mass, but higher ratios at low masses. The relationship between black hole and stellar mass will be discussed in more detail in Section \ref{stellarbh}.

The higher $\epsilon_w$ runs also show reduced star formation rates around $M_*=10^{10} M_\odot$. This suggests that the strong wind activity can play an active role in quenching smaller galaxies. 

\subsubsection{Removing the X-ray Mode}

One element of RAFIKI that differs from SIMBA is the lack of an X-ray feedback mode. This mode was meant to emulate heating caused by X-rays from the accretion disk, and was activated in the most massive galaxies when the jet mode was also on. In SIMBA, it was found that without this X-ray mode, it was difficult to fully quench the most massive galaxies, as the upper end of the GSMF was overpredicted and the sSFR was slightly higher in these massive systems. As shown in Figure \ref{fig:gsmf}, while there may be a slight overprediction of massive galaxies at $z=0$ for $\epsilon_j = 0.1$ and 0.2, the more powerful $\epsilon_j = 0.3$ runs align well with the data at the high mass end. 

The stellar mass- specific star formation rate (sSFR) can be used to probe in more detail the impacts of removing the X-ray mode. As shown in Figure \ref{fig:z0_stellar_sfr}, systems with log($M_*/M_\odot) > 11$ show an uptick in their sSFR. This is most apparent for lower $\epsilon_j$ values,  as at $\epsilon_j = 0.3$ the sSFR of galaxies with log($M_*/M_\odot) \approx 12$ are an order of magnitude lower than those run with $\epsilon_j = 0.1$. Thus we see that stronger jets are able to quench massive systems without the need for additional feedback modes. These results appear to be independent of $\epsilon_w$, again pointing to the wind mode playing little to no role in the most massive systems. Additionally, while $\epsilon_j = 0.1$ is still sufficient to reproduce the high-mass GSMF, it might not quite be strong enough to realistically quench the highest-mass systems.

\subsection{Connections between Stars, Halos, and Supermassive Black Holes}
 
\begin{figure*}
\includegraphics[width=\linewidth]{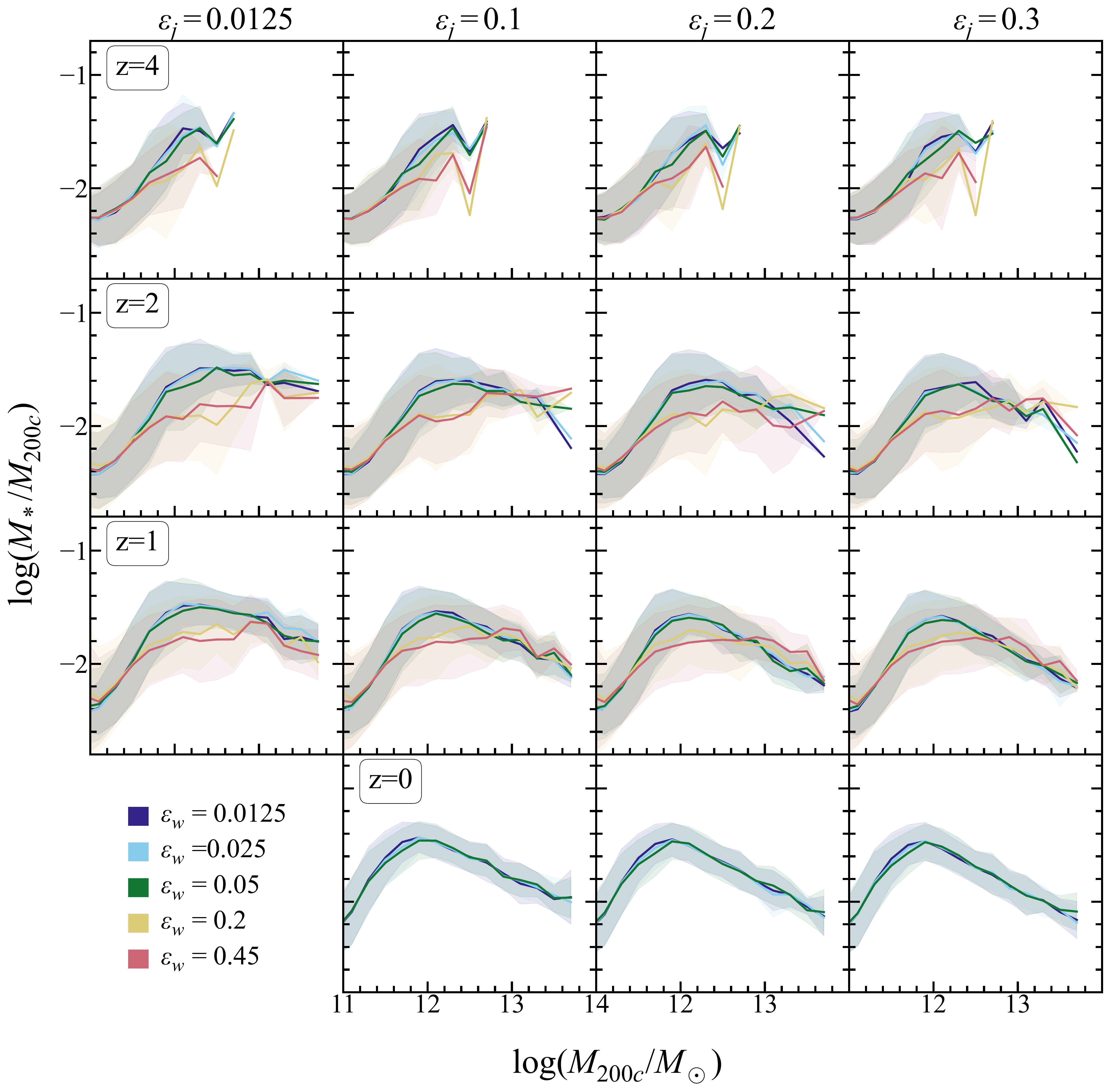}
\caption{Stellar-halo mass ratio as a function of halo mass for central galaxies at $z$=4,2,1, and 0 (top to bottom). Each column represents a different value for $\epsilon_j$, with colors correspond to $\epsilon_w$. Here we see how boosting the efficiency of the wind phase prevents the stellar growth of galaxies in halos with $M_{200c} \approx 10^{12.5}$, while leading to larger stellar mass galaxies in low-mass halos.}
\label{fig:stellar-halo}
\end{figure*}

\subsubsection{Stellar Mass: Halo Mass Relationship}

Understanding the relationship between stellar and halo masses is key to studies of galaxy formation. In massive systems, dark matter dominates the gravitational potential, shaping gas heating and cooling, while feedback suppresses star formation, especially in large halos.

Figure \ref{fig:stellar-halo} shows the stellar-halo mass ratio (hereafter SHMR) as a function of halo mass for all RAFIKI runs. The SHMR is parameterized by a turnover at a characteristic mass that decreases with redshift to $M_{200} \approx 10^{11.75} M_\odot$ at $z=0$. In the RAFIKI runs, we see this shape clearly for $\epsilon_w$ values of 0.0125-0.05. At higher wind efficiencies, however, we see a suppression of the SHMR at the characteristic mass and a slightly increased value past $M_{200} \approx 10^{13} M_\odot$, leading to a relatively flat profile above the characteristic mass. The stronger winds are leading to less quenching at high mass scales, most likely due to the removal of accreting material from around the central SMBH. These trends also suggest that the more efficient winds are over-quenching galaxies around the characteristic halo mass. This aligns with the trends seen in Figure \ref{fig:main_sequence}, where runs with $\epsilon_w = 0.45$ generate low SFR in galaxies with $M_* \approx 10^{10} M_\odot$. As the behavior of the SHMR in the high wind efficiency runs are inconsistent with observed trends \citep{Girelli2020}, runs with $\epsilon_w = 0.2$  and 0.45 were not continued to $z=0$. 

\subsubsection{Bulge Mass - Black Hole Mass Relationship} \label{stellarbh}
\begin{figure*}
\includegraphics[width=\linewidth]{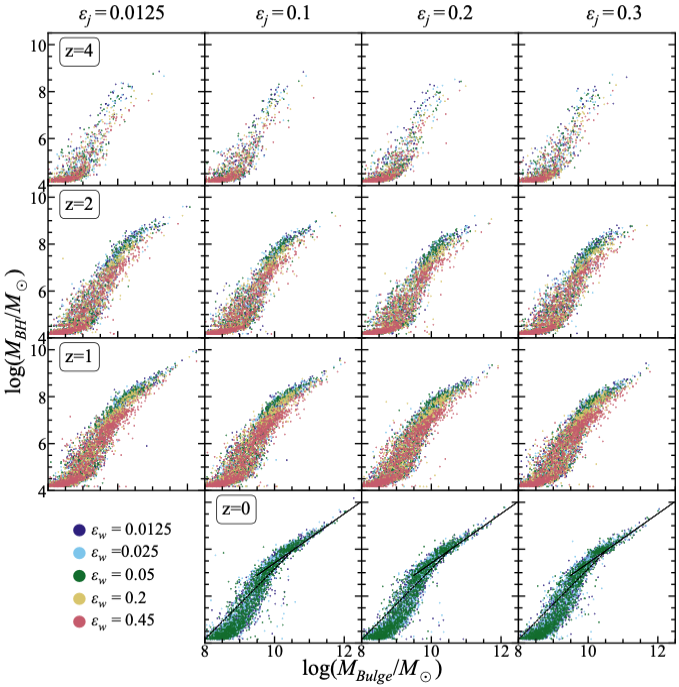}
\caption{Central galaxy bulge-black hole mass relationship at redshifts 4,2, 1, and 0 (top to bottom). Each column represents a different value for $\epsilon_j$, with jet mass loadings at 2.5, 20, 40, and 60 $\%$ from left to right. At z=0, the solid black line represents the best fit power law from \citep{mcconnell2013}, and the dashed black line gives the best fit for lower mass galaxies taken from \citep{bentz2018}. }
\label{fig:stellar-bh}
\end{figure*}

The relationship between the bulge and black hole mass in galaxies motivates a strong connection between galaxy stellar properties and their central supermassive black holes. In simulations, this relationship is usually governed by the radiative efficiency, $\eta$, set to 0.1 for all RAFIKI runs. Figure \ref{fig:stellar-bh} shows the stellar bulge-black hole mass relationship for all RAFIKI runs, where the bulge mass is defined within CAESAR as twice the counter-rotating mass. At higher wind efficiencies, we see that black hole growth is limited. By a redshift of one, the run with $\epsilon_w =0.45$ has very few galaxies with $M_{\rm BH}>10^{7.5} M_\odot$. With $\epsilon_w=0.2$, there are more massive black holes, but still not as many as the runs with lower mass loadings. We see almost no change in the stellar-black hole mass relationship for different $\epsilon_j$ values. 

This variation with $\epsilon_w$ but not $\epsilon_j$ highlights that an isotropic wind feedback mechanism plays a crucial role in regulating black hole growth, far more than high velocity bipolar jets. 

Figure \ref{fig:stellar-bh} also shows the best fit relationship from observations of galaxies in \citep{mcconnell2013, bentz2018}. In general, the RAFIKI runs match this relationship, perhaps slightly overpredicting black hole masses in high mass galaxies.


\begin{figure*}
\includegraphics[width=\linewidth]{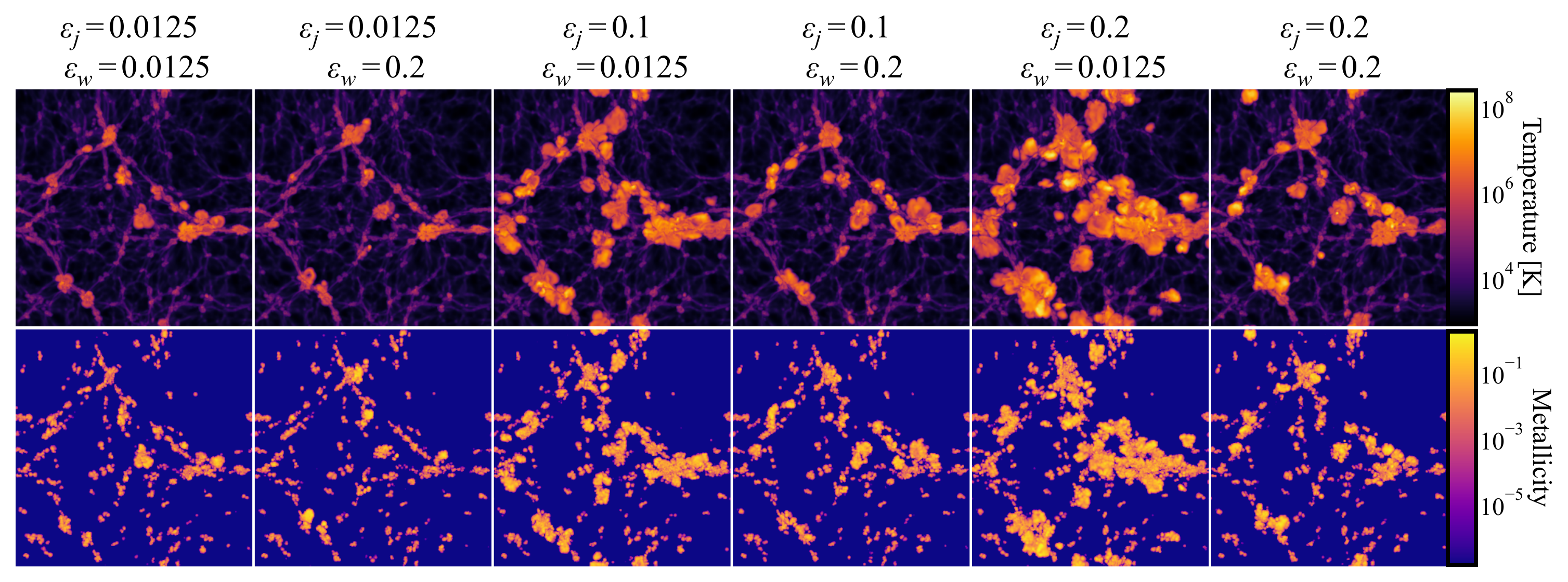}
\caption{Slice plots showing the temperature (top), and metallicty in solar units (bottom) for six RAFIKI runs at $z$=1. The two leftmost columns show runs with $\epsilon_j = 0.0125$, and the two rightmost show  $\epsilon_j = 0.2$, demonstrating the large impact jets play in shape gas properties on large scales. We see high $\epsilon_j$ values increasing the radius at which heat and metals are distributed, while high $\epsilon_w$ values actually decreases the radius.}
\label{fig:z1_images}
\end{figure*}

\begin{figure*}
\includegraphics[width=\linewidth]{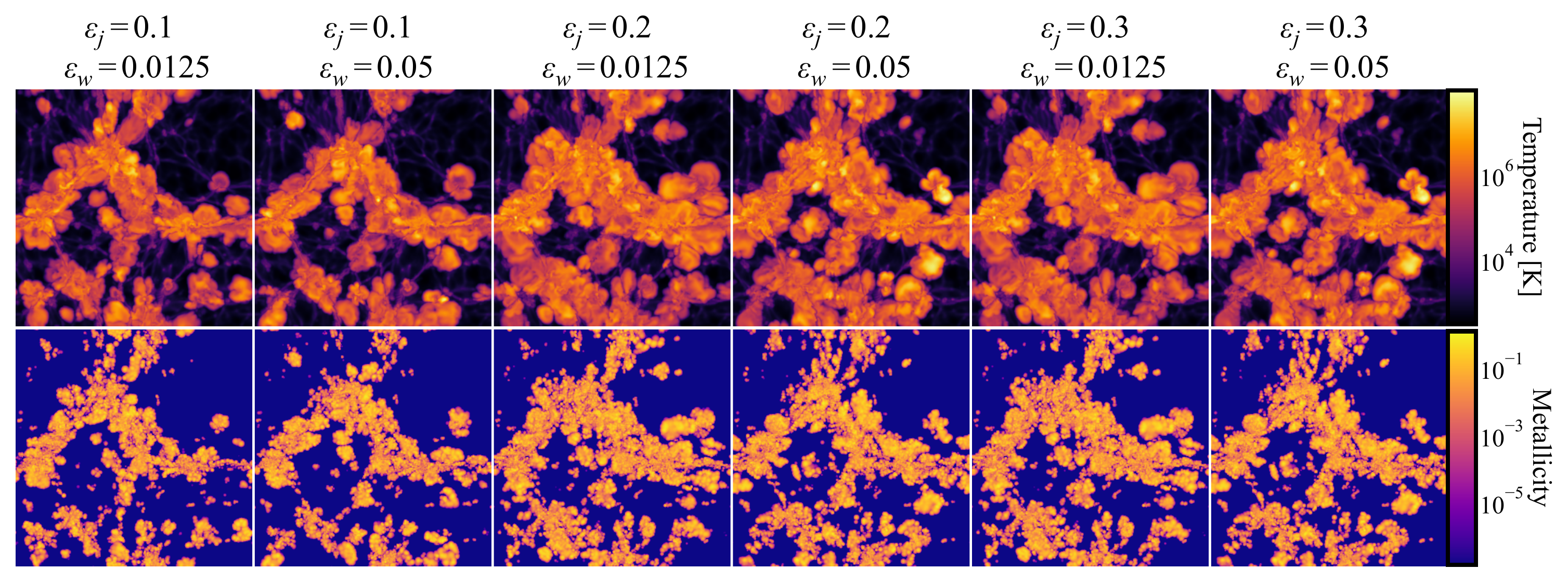}
\caption{Slice plots showing the temperature (top), and metallicty in solar units (bottom) for six RAFIKI runs at $z$=0. The two leftmost columns show runs with $\epsilon_j = 0.1$, middle two columns show $\epsilon_j = 0.2$, and the two rightmost show  $\epsilon_j = 0.3$. All six runs shown here generally match global properties well, yet produce different CGM conditions. We see that these runs do result in different baryon properties, despite all succesfully matching galaxy global properties. }
\label{fig:z0_images}
\end{figure*}
\subsection{Gas Properties}

\subsubsection{Overall Distribution}
Understanding the effects of AGN feedback on baryons is crucial for developing a comprehensive theory of galaxy formation. The wide range of parameter space covered by RAFIKI provides an exciting opportunity to explore the conditions of the galactic environment and the baryon cycles that shape galaxies. 

Figure \ref{fig:z1_images} shows slice plots of gas properties in selected RAFIKI runs at $z$=1. Here we see the extent to which jets impact the gas distribution, heating and distributing metals out to significant radii. Elevating the wind efficiency compared to fiducial SIMBA does not help distribute metals or heat further from the galaxy.

For the RAFIKI runs ran to $z$=0 that successfully quench galaxies, there is less variation in the CGM, although there are some significant differences. Figure \ref{fig:z0_images} shows slices of the gas properties in six runs at $z$=0. While the most obvious differences again come from changing $\epsilon_j$, variations in $\epsilon_w$ also have an impact. Qualitatively, increased $\epsilon_w$ leads to less far reaching impact, with heightened temperature and metallicity structures being less extended. However, when $\epsilon_j=0.3$ we no longer see this trend, with a puffier structure in the high $\epsilon_w$ run compare to the low.

\subsubsection{Gas Fractions}
\begin{figure*}
\includegraphics[width=\linewidth]{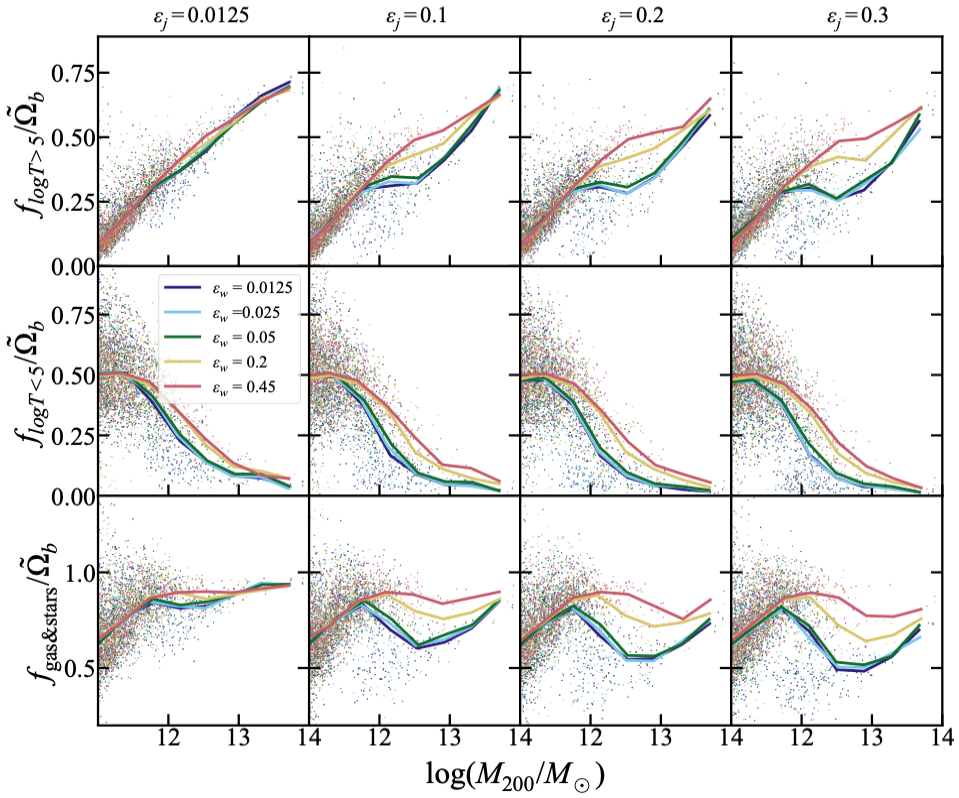}
\caption{Gas fractions as a function of $M_{200}$ at  $z=1.$ From top to bottom, each row shows hot gas ($T>10^5$), cool gas ($T<10^5$), and all baryon fractions. Each column represents a different value for $\epsilon_j$, with colors correspond to $\epsilon_w$. The solid lines show the running median for each run. We see that both $\epsilon_j$ and $\epsilon_w$ impact the gas fractions across a range of halo mass scales.}
\label{fig:gas_frac}
\end{figure*}

We can start to quantify the baryonic differences of RAFIKI runs by looking at the halo gas fractions. AGN feedback significantly shapes the distribution of gas around halos, both by ejecting material to larger radii and by depositing heat that can unbind the gas from the halo. Thus the baryon contents of halos can differ significantly from the mean cosmic fraction. Figure \ref{fig:gas_frac} shows the fraction of for three phases: hot gas ($T>10^5$), cool gas ($T<10^5$), and all gas and stars relative to $M_{200}$. The values are scaled to $\tilde{\Omega}_b=\Omega_b/\Omega_M = 0.16$, such that halos with a fraction of 1 have their cosmic share of baryons. The lines show the running median for ten halo mass bins. 

We see that both $\epsilon_j$ and $\epsilon_w$ have significant impacts on baryon fractions. In examining the hot gas and the total baryon budget, with $\epsilon_j= 0.0125$ we do not see the dip in fractions at $M_{200} \approx 10^{12}$ visible at higher $\epsilon_j$. This highlights the key role an ejective feedback mechanism can play at the group scale. The jets also have a significant impact at cluster scales ( $M_{200} > 10^{13.5}$). At these masses and low $\epsilon_w$ values, $\epsilon_j= 0.0125$ results in a total baryon fraction of 0.95, compared to 0.7 for $\epsilon_j= 0.3$. This difference decreases with increasing $\epsilon_w$, but demonstrates that increasing the efficiency of an ejective model does serve to remove baryons even in systems with a high gravitational potential. 

Higher $\epsilon_w$ values result in larger gas fractions, a trend seen across all phases. As the differences increase with $\epsilon_j$, this could point more to increased wind efficiencies preventing jets from turning on, as opposed to the effects of the wind mode itself. It is  worth noting that there is very little change in baryon fractions from $\epsilon_w = 0.0125$ to 0.05.

\begin{figure}
\includegraphics[width=\linewidth]{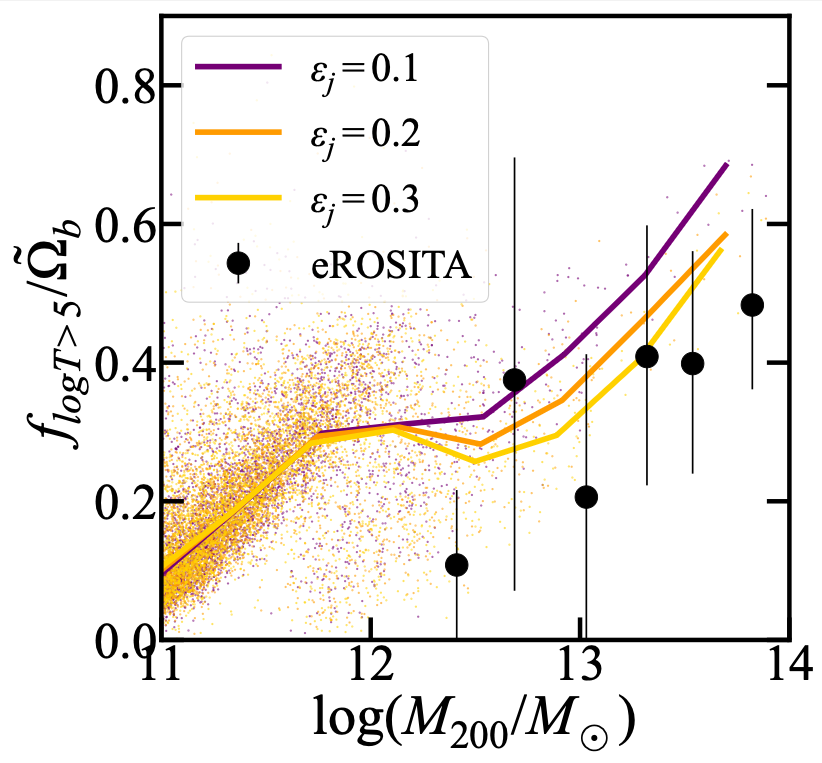}
\caption{Hot gas fractions as a function of $M_{200}$ at $z=0$ for three RAFIKI runs with $\epsilon_w = 0.0125$. The solid lines show the running median for different choices of $\epsilon_j$. Black points show observations from \cite{popesso2024}, highlighting a preference for higher feedback efficiencies in order to match X-ray data.}
\label{fig:gas_frac_z0}
\end{figure}

Despite the minimal effects of the wind mode, the impacts of $\epsilon_j$ make the RAFIKI suite a promising tool for constraining the impacts of AGN feedback on baryon fractions. Figure \ref{fig:gas_frac_z0} shows the hot gas fraction calculated within $R_{200}$ for three RAFIKI runs at $z=0,$ compared to the latest group-scale constraints from eROSITA \citep{popesso2024}. eROSITA results have been suggesting lower gas fractions on the group and cluster scale than previously found, leading to inconsistencies with several leading simulations \citep{popesso2024}. The RAFIKI results, however, are generally consistent with observations, particularly for $\epsilon_j=0.3$. This favors strong ejective feedback in massive systems, although as seen in Figure \ref{fig:sfrd} this can lead to inconsistencies with global stellar properties. 

\section{Discussion}\label{sec3}

In this work, we set out to understand if the low-velocity wind feedback model implemented in SIMBA-C can quench massive galaxies if the mass loadings are set separately from the jet mode. We modified the kinetic prescriptions of AGN feedback in to be parameterized by two efficiency values: $\epsilon_w$ for the wind model, launched isotropically, and $\epsilon_j$ for the higher-velocity collimated outflows. The resulting RAFIKI suite consists of twenty runs that vary these two parameters, allowing for a systematic comparison with a range of galaxy property distributions and scaling relations.

From the galaxy stellar mass function, we see that quenching in high mass galaxies requires significant kinetic power in the higher velocity jets, even in models with significantly elevated energies in the wind mode. This demonstrates a key element of how quenching functions in SIMBA-C. In high mass halos, the wind mode does not reach adequate velocities to escape the gravitational potential, constraining its effects to small scales. Yet even with significantly higher energy input within the galaxy, quenching is unable occur. This points to the quenching mechanism in SIMBA-C being reliant on large scale heating of the circumgalactic medium, and the ejection of gas to large radii, as opposed to a more internal process shaped by heating of the interstellar medium. Work done in \cite{hopkins2016} with smaller scale simulations has also found that winds have little effect outside of the area they are immediately evacuating, supporting the idea that without higher velocities this wind mode has limited galaxy-scale effects. They also found that changing the momentum flux did not change these results, aside from impacting the volume of the evacuated area around the black hole. This aligns with the lack of large scale differences we see in changing the mass-loading of the winds in RAFIKI. The reliance on large-scale energy deposition has also been found on cluster scales, with \cite{husko2024} exploring the effects of jet and wind modes in simulated idealized clusters and finding that jets were needed to quench the cooling flow around central galaxies. It is notable that we are seeing similar trends on smaller mass scales in the RAFIKI suite. 

While the wind mode is thus shown to not quench high mass galaxies, the RAFIKI suite also allows for a more in-depth exploration of the impact winds can have at lower mass ranges. We find that winds have three major effects:
\begin{itemize}
\item They play a key role in regulating black hole growth, something jets do not accomplish, potentially because their energy is deposited too far away. This leads to reduced quenching in massive systems, as with constricted black hole growth, AGN feedback in the jet mode is less energetic.
\item They directly impact star formation in intermediate mass galaxies ($M_* \approx 10^{10} M_\odot$), and they indirectly affect higher-mass systems. At very high $\epsilon_w$ values, this can result in significant disruption of the stellar-halo mass ratio, flattening the profile at high masses. 
\item They generate some variation in CGM conditions around massive galaxies. While the winds cannot achieve the velocities necessary to deposit material at large radii, the interplay between the wind and jet modes means that strengthening the winds serves to weaken the jets. This in turn impacts the distribution of metals and energy in the CGM. 
\end{itemize}

Beyond the effects of the wind mode, the RAFIKI suite has also revealed a degeneracy within the jet efficiency parameters. Values spanning a factor of three all adequately match the galaxy properties. However, the different values change the baryon distribution and the extent to which AGN impact temperatures and metallicities in the CGM. This makes RAFIKI a useful tool for future studies motivated by understanding how feedback shapes the underlying matter distribution and CGM conditions.

It is worth noting that the runs that successfully match the galaxy stellar mass function do not necessarily provide the best fit for other galaxy properties. The star formation rate density is underpredicted by runs with high $\epsilon_j$, as is the relationship between stellar mass and star formation rate. It also possible that at the very highest mass ranges the RAFIKI parameters are not able to fully quench galaxies, although given the small sample sizes at these scales, larger boxes will be needed to understand high mass halos. 

Overall, the results presented here highlight a fairly large degenerate parameter space for kinetic models of AGN feedback. Runs with $0.1<\epsilon_j<0.3$ are all generally successful at quenching massive systems, and $0.0125<\epsilon_w<0.05$ is a similarly degenerate region in the parameter space. Further constraining these parameters will require detailed observations of baryons, as the conditions in the circumgalactic medium and gas fractions change significantly within this parameter space, especially when varying $\epsilon_j$. As the wind mode has minimal impact on these larger scales, further work is needed exploring smaller scale systems in higher resolution boxes where the effects of the wind mode can be more realistically modeled.

 RAFIKI serves as a valuable addition to modern cosmological simulations and is well suited for further study that seeks to constrain AGN feedback models. Spanning a range of parameter space within a modified SIMBA-C model allows for a focused study on the role specific feedback modes play. The cosmological volume allows for explorations across mass scales and cosmic times. Future runs would benefit from exploring the velocity parameter space in more detail. Currently, the wind mode velocities do not exceed $\approx$ 1000 km/s, but many recent observations show that radiation-driven winds can reach much higher velocities\citep[e.g.][]{Zakamska2016,fiore2017, xu2025}. There has been several recent attempts to model these faster outflows in smaller scale, higher-resolution simulations. \cite{costa2020} modelled winds of speeds ranging from 5000 to 30000 km$s^{-1}$, finding those winds could power galactic scale outflows with some effect on star formation, although they do not cause complete and rapid quenching. Faster velocities could also effect black hole growth and scaling relations, as \cite{hopkins2016} found winds with velocities of 30000 km$s^{-1}$ did not blow out as much of a bubble in the circumnuclear gas, allowing for higher accretion rates than lower velocity models.  This work focused on modifying the mass loading element of the kinetic power in order to isolate the effects of that single feedback parameter, but incorporating a wider range of velocities would help develop a more complete physically motivated sub-grid wind model.

 As is, there are many avenues for future work using the RAFIKI suite, including exploring connections with other processes such as stellar feedback and examining the multi-phase structure of the gas more closely. The RAFIKI suite will also be used to generate predictions for multiwavelength CGM surveys, assessing the observability of the different conditions generated by the range of models. 

\section*{Acknowledgements}
We would like to thank Seth Cohen, Darby Kramer, Brad Koplitz, and Phil Mauskopf for their helpful discussions during the development of this work. We also thank Robert Thompson for developing CAESAR, and the yt team for the development and support of this useful community tool. S.G. acknowledges support from NSF Grant No. 2233001. E.S. acknowledges support from NASA grants 80NSSC22K1265 and 80NSSC25K7299. AB acknowledges funding support from the Infosys endowment at IISc, the Leverhulme Trust, and Natural Sciences and Engineering Research Council of Canada. AB would like to thank the IfA (Univ. of Edinburgh) for hospitality during his recent visits.  
\vspace{5mm}

\newpage

\bibliographystyle{aasjournal}
\bibliography{bib.bib}

\end{document}